\documentclass[manuscript]{acmart}
\usepackage{float}
\usepackage[prependcaption,textsize=tiny]{todonotes}
\usepackage{amsthm}
\usepackage{mathtools}
\usepackage{csquotes}
\newtheorem{theorem}{Theorem}[section]

\newtheorem{prop}[theorem]{Proposition}
\newtheorem*{claim}{Claim}

\newcommand{\R}{\mathbb{R}}
\newcommand{\N}{\mathbb{N}}
\newcommand{\E}{\mathbb{E}}

\newcommand{\surprise}{{\texttt{surp}}}
\newcommand{\sgn}{\texttt{sgn}}
\newcommand{\hist}{\texttt{Hist}}
\newcommand{\conv}{\operatorname{conv}}
\renewcommand{\P}{\mathbb{P}}
\newcommand{\ME}{\operatorname{ME}}
\newcommand{\OC}{\operatorname{OC}}
\newcommand{\HA}{\operatorname{HA}}
\newcommand{\uniform}{\texttt{uniform}}
\newcommand{\softmax}{\texttt{softmax}}

\newcommand{\Persistent}{\texttt{Persistent softmax}}
\newcommand{\constant}{\texttt{constant}}
\newcommand{\greedy}{\texttt{greedy}}

\usepackage{placeins}
\usepackage{subcaption}
\usepackage{dsfont}

\copyrightyear{2022} 
\acmYear{2022} 
\setcopyright{rightsretained} 
\acmConference[RecSys '22]{Sixteenth ACM Conference on Recommender Systems}{September 18--23, 2022}{Seattle, WA, USA}
\acmBooktitle{Sixteenth ACM Conference on Recommender Systems (RecSys '22), September 18--23, 2022, Seattle, WA, USA}
\acmDOI{10.1145/3523227.3546778}
\acmISBN{978-1-4503-9278-5/22/09}

\acmConference[ACM RecSys '22]{ACM Conference on Recommendation Systems}{September 18-23, 2022}{Seattle, U.S.A.}
\begin{document}
% \footnotetext{Authors have contributed equally.}
% \renewcommand{\thefootnote}{*}

\title{Towards Psychologically-Grounded Dynamic Preference Models}
\author{Mihaela Curmei$^\ast$}
\email{mcurmei@berkeley.edu}
\affiliation{%
  \institution{University of California, Berkeley}
  \streetaddress{253 Cory Hall}
  \city{Berkeley, CA}
  \country{USA}}
\author{Andreas Haupt$^\ast$}\thanks{$^\ast$ Equal contribution. $^\dagger$ Equal advising contribution.}
\email{haupt@mit.edu}
\affiliation{%
  \institution{Massachusetts Institute of Technology}
  \streetaddress{32 Vassar Street}
  \city{Cambridge, MA}
  \country{USA}}
 \author{Dylan Hadfield-Menell$^\dagger$}
\email{dhm@csail.mit.edu}
\affiliation{%
  \institution{Massachusetts Institute of Technology}
  \streetaddress{32 Vassar Street}
  \city{Cambridge, MA}
  \country{USA}}
\author{Benjamin Recht$^\dagger$}
\email{brecht@berkeley.edu}
\affiliation{%
  \institution{University of California, Berkeley}
  \streetaddress{253 Cory Hall}
  \city{Berkeley, CA}
  \country{USA}}

\begin{abstract}
Designing recommendation systems that serve content aligned with time varying preferences requires proper accounting of the feedback effects of recommendations on human behavior and psychological condition. We argue that modeling the influence of recommendations on people's preferences must be grounded in psychologically plausible models. We contribute a methodology for developing grounded dynamic preference models. We demonstrate this method with models that capture three classic effects from the psychology literature: Mere-Exposure, Operant Conditioning, and Hedonic Adaptation. We conduct simulation-based studies to show that the psychological models manifest distinct behaviors that can inform system design. Our study has two direct implications for dynamic user modeling in recommendation systems. First, the methodology we outline is broadly applicable for psychologically grounding dynamic preference models. It allows us to critique recent contributions based on their limited discussion of psychological foundation and their implausible predictions. Second, we discuss implications of dynamic preference models for recommendation systems evaluation and design. In an example, we show that engagement and diversity metrics may be unable to capture desirable recommendation system performance.
\end{abstract}
\maketitle
 
%%%%%%% INTRODUCTION.  %%%%%%%%%
\section{Introduction}
% Recommendation system mediate interactions between people and content in the digital world. They do so with the aim to surface content that will be consumed and align with individuals' preferences.

% In many current recommendation systems, preferences are implicitly or explicitly assumed to be static throughout time and not altered by recommendation. There are several reasons that point to a need for modeling of how recommendation systems affect user preferences. In evaluation, for example, several (mostly empirical) studies have observed that offline metrics for recommendation systems performance are not indicative of online performance \cite{rossetti2016contrasting,garcin2014offline,jeunen2019revisiting,krauth2020offline}. A potential source of the mismatch is the influence that recommendation have on user preferences. In the study of societal impacts of recommendation, several models \cite{rossi2018closed, jiang2019degenerate, geschke2019triple} study the role of recommendation in creating polarization, filter bubbles, and extremism. The conclusions of these models rely on modeling choices of how reactions happen, and need to make realistic assumptions on user preferences.

Recommendation systems mediate interactions between people and content in the digital world. They attempt to surface content that will be consumed and aligns with individuals' preferences.

In much of recommendation systems research, preferences are implicitly or explicitly assumed to be static throughout time and not altered by recommendation. While useful in many contexts, this assumption has drawbacks. For example, static preferences in user models may account for the poor generalization performance of offline metrics in predicting deployed recommendation performance \cite{rossetti2016contrasting,garcin2014offline,jeunen2019revisiting,krauth2020offline}. Furthermore, dynamic preferences have been central to arguments of negative impacts of recommendation, such as polarization, filter bubbles, and extremism \cite{rossi2018closed, jiang2019degenerate, geschke2019triple, chaney2018algorithmic}.

This study considers the design and validation of dynamic user preference models. We propose that behavioral models should be designed with, at least, two desiderata. 

First, they should be grounded in experimental psychological evidence. A wealth of research in experimental psychology identifies several such behavioral patterns, or \emph{psychological effects}. We contribute a methodology to leverage these results and design psychologically plausible dynamic preference models. We demonstrate this method by designing models to capture three established findings about preference formation and shift  in humans: \emph{Mere Exposure}, \cite{hekkert2013mere,nordhielm2002influence, fang2007examination,cox2002beyond}, where simply experiencing a stimulus (e.g., a piece of content) tends to make humans view it more positively; \emph{Operant Conditioning} 
\cite{fagerstrom2011study,helson1947adaptation,foxall2017behavioral}, where preferences shift towards actions that are associated to \enquote{positive feedback} (e.g., consuming \enquote{better than expected} content); and \emph{Hedonic Adaptation} \cite{chugani2015happily,nelson2008interrupted,yang2015sentimental}, where  satisfaction levels (e.g., with content) return to a baseline after a period of time. 

Second, models should be verified in simulation for plausibility. For any given effect, there are many ways it could be formalized mathematically. Simulations of user-recommender dynamics can help with these modeling choices. Preference representations should produce plausible predictions across a range of recommender system designs and initial conditions. For example, a model where behaviour shifts to consuming a single type of content with probability close to one is unlikely to match real observations.

\begin{figure*}
    \centering
    \includegraphics[width=0.8\textwidth]{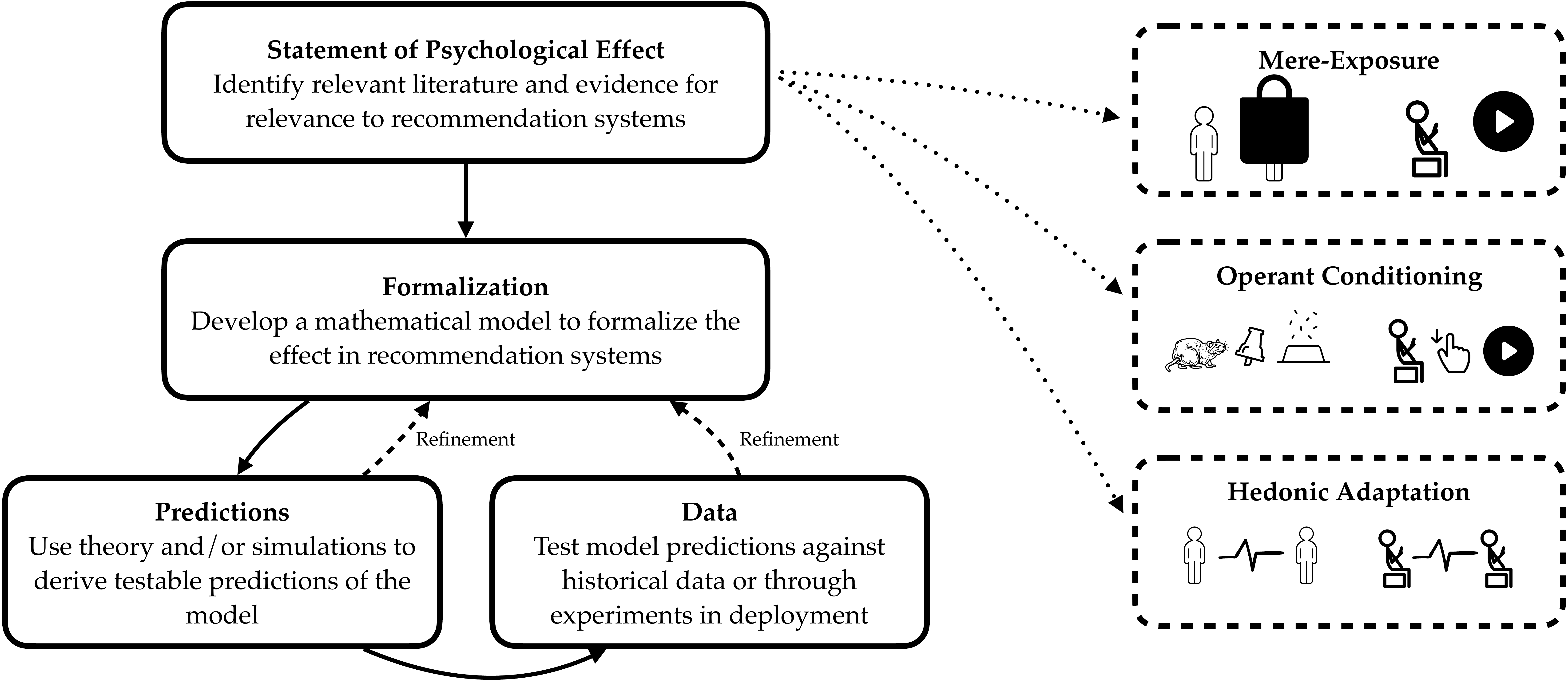}
    % \begin{subfigure}[b]{.3\linewidth}
    % \includegraphics[width=0.85\linewidth]{figures/zatonc.png}
    % \caption{Mere-Exposure, in \cite{zajonc1968attitudinal} and in\newline Recommendation Systems}
    % \label{subfig:zajonc}
    % \end{subfigure}
    % \begin{subfigure}[b]{.3\linewidth}
    % \includegraphics[width=0.85\linewidth]{figures/skinner.png}
    % \caption{Operant Conditioning, in \cite{skinner2019behavior} \newline and in Recommendation Systems}
    % \label{subfig:skinner}
    % \end{subfigure}
    % \begin{subfigure}[b]{.3\linewidth}
    % \includegraphics[width=0.85\linewidth]{figures/brinkmann.png}
    % \caption{Hedonic Adaptation, in \cite{brickman1971hedonic} and in \newline Recommendation Systems}
    % \label{subfig:brickman}
    % \end{subfigure}
    \caption{The methodology derived in this article. First, formulate a psychological theory that should ground a particular effect. Then, formalize it, and discipline it with predictions derived from simulations and data from deployed recommendation systems. This article exemplifies this methodology for three effects: Mere-Exposure Effect, which we introduce using Goetzinger's \cite{zajonc1968attitudinal} classical experiment, Operant Conditioning, which we introduce using Skinner's experiments \cite{skinner2019behavior}, and hedonic adaptation, which we introduce using Brickman's study of happiness after extreme life events \cite{brickman1971hedonic}. The classical experiments are depicted in vignettes on the right.}
    \label{fig:vignettes}
\end{figure*}

%There may be many mathematical models that seem to capture each psychological effects. Further selection of models is possible by studying predictions of mathematical models in combination with standard recommenders, and their predictions. Some predictions might not be plausible, or not align with observations that a recommendation system designer has. Having a combination of predictions of mathematical modeling and background in psychology helps to discuss models both for recommendation system evaluation and design, as well as the theoretical study of societal impacts of recommendation. 

%After psychological grounding of models, there might still be several mathematical models that reasonably match a psychological effect. To critique mathematical models, simulations of user-recommender dynamics with standard recommenders are promising tools. Some predictions obtained from simulations might not be plausible, or not align with other evidence accessible to a recommendation system designer. 
Having a combination of predictions from simulations and background from psychology feeds back into improving user modeling and hence recommendation system evaluation and design. This may help address the on-/offline gap in evaluation as well as increase the credibility of the theoretical study of societal impacts of recommendation.

% The contributions of this article  are threefold:
% \begin{enumerate}
%     \item We propose a broadly applicable methodology for psychologically grounding dynamic preference models;
%     \item We exemplify this methodology, by considering three effects studied in the psychology literature. We review literature on these effects with a focus on relevance for recommendation systems, propose formalizations of the effects, and, in simulations, characterize qualitative behaviors and derive testable predictions of user-recommender systems using formalized effects;
%     \item Finally, we discuss two implications of this study for recommendation system evaluation and design. First, we critique recent contributions based on their limited discussion of psychological foundations and their implausible predictions. Second, we demonstrate in an example that recommendation system metrics such as engagement and diversity are unable to capture desirable recommendation system performance in the presence of dynamic user preferences.
% \end{enumerate}

The contributions of this article are threefold:
\begin{enumerate}
\item We consider three case studies of behavioral effects due to consumption established in psychological research. We review the relevant literature of these effects with a focus on their applicability to recommendation systems. We propose a modeling framework to mathematically formalize these effects. Finally, using simulations, we characterize qualitative properties of our proposed models and derive testable predictions that can be used to test the validity of a particular mathematical formalization independently of the applicability of a behavioral effect.\footnote{While the models we present are classical in psychology, we do not claim that these are models will fit empirical data in recommendation systems well. Our mathematical formalization of theories is particularly parsimonious by formulating preference shifts as changes to vector of user factors. This parsimony, however, leads to simplifications of several potentially psychologically relevant factors, which we do not claim to exhaustively model. First, our models do not differentiate between exposure to and consumption of content. Furthermore, they assume that preferences are purely evolving instead of formed, i.e. users have preferences for even unseen content. In addition, the models are formulated abstractly, as opposed to for a concrete application domain, suppressing features that might make different psychological effects particularly salient in different domains. Our models are exemplary for the application of a methodology, and will need empirical validation before used to model interactions in a real systems. The selection of good models, we claim, however, benefits from inclusion of psychological grounding and extensive testing.} 
\item We synthesize the case studies into a broadly applicable methodology for psychologically grounding dynamic preference models. We propose a multi-step procedure which first calls for explicitly stating and providing evidence in support of a psychological effect. Following this one iterates by proposing a mathematical formalization for the behavioral model, deriving testable predictions and comparing them against data relevant to the recommendation context. The proposed methodology is depicted in \autoref{fig:vignettes}.
\item Finally, we discuss two implications of this study for recommendation system evaluation and design. First, we critique recent contributions based on their limited discussion of psychological foundations and their implausible predictions. Second, we demonstrate in an example that recommendation system metrics such as engagement and diversity are unable to capture desirable recommendation system performance in the presence of dynamic user preferences.
\end{enumerate}

\subsection{Outline}\label{sec:outline}
The rest of this article is structured as follows. In \autoref{sec:litrev}, we position our work in the related literature. We first derive and exemplify the methodology proposed in this article. In \autoref{sec:examples}, we consider three psychological effects, and review psychological evidence most relevant to recommendation systems contexts. We mathematically formalize these three psychological effects in \autoref{sec:setup}. We present experimental results in \autoref{sec:predictions}. Section \ref{sec:discussions} contains discussion. In \autoref{sec:plausibility}, we formalize the methodology used in this study and discuss how its application may help to refine existing models in the recent literature. In \autoref{sec:design}, we illustrate with an example the brittleness of commonly used metrics in the design and evaluation of recommendation systems. We collect avenues for future work in \autoref{sec:conclusions}. Proofs and additional simulations can be found in the supplementary materials. Code to reproduce our simulations can be found in the supplemental materials.

%%%%%% RELATED WORK.  %%%%%%%%%
\section{Related Work}\label{sec:litrev} %Ready for MC
%This paper relates to several branches of literature on recommendation systems
%where recommendation induced shifts in preferences and behaviors of users are a central object of study.
\paragraph{Evaluation}
First we relate to literature of recommendation system evaluation. Preference changes in response to recommendation may affect the evaluation of recommendation systems. The main design paradigm of recommendation systems is to use offline train-test splits to design and select recommendations. Several papers argue that offline metrics are poor indicators of online performance \cite{jeunen2019revisiting,garcin2014offline,beel2013comparative,sun2020we}. Often the lack of external validity of current evaluation methodology is attributed to unmodeled feedback dynamics between users and recommenders \cite{rossetti2016contrasting, krauth2020offline}, which further motivates the need to study dynamic preference models.
\paragraph{Societal Impacts}
Well-founded dynamic preference models can help resolve apparent contradictions among findings on the societal impacts of recommendation systems. On one side, several studies claim that dynamic interaction of recommendation systems with users may lead to polarization \cite{dandekar2013biased, rossi2018closed}, filter bubbles \cite{ pariser2011filter,geschke2019triple}, homogenization  \cite{chaney2018algorithmic}, echo chambers \cite{noordeh2020echo, donkers2021dual, jiang2021mechanisms} and extremism \cite{munn2019alt, ribeiro2020auditing, jiang2019degenerate}. Empirical audits, however, find that algorithmically recommended content is more diverse than natural consumption \cite{nguyen2014exploring}, and that real systems do not exhibit the strong extremism or polarization effects implied by theoretical models \cite{markmann2021youtube, ledwich2019algorithmic, hosseinmardi2020evaluating, tomlein2021audit, levy2021social}. %Other research argues that even though over time online media spaces tend to exhibit undesirable dynamics, there is not enough evidence to attribute them to algorithmic manipulation \cite{moller2018not}. 
The study \cite{carroll2021estimating} points out that recommendation systems might lead to undesired changes in preferences, and proposes to design for \emph{safe preference shifts}, which are preference trajectories that are deemed \enquote{desirable}.
%\paragraph{Mitigation of Recommendation Biases} Beyond the study of societal impacts, also algorithmic interventions depend on accurate dynamic user models, and can inform re-ranking \cite{pei2019personalized,geyik2019fairness,abdollahpouri2019managing} that mostly tackle diversity and fairness goals, but relies on prediction of how users react to content recommendation, variation of the level of exploration \cite{jadidinejad2020using, song2021show} and type of exploration \cite{khenissi2020theoretical} which aim to mitigate popularity bias.
\paragraph{Dynamic Preference Models}
 Some of the existing dynamic preference models assume that the preferences change independent of the recommendations. Examples are Dynamic Poisson Factorization (DPF) \cite{charlin2015dynamic, hosseini2018recurrent}, which assumes  that users and content items have latent representations which evolve as Gaussian random walks; and \cite{koren2009collaborative}, which models behavioral changes as external concept drifts which affect item quality and average user rating levels. Other works consider the feedback loop between recommendation systems and users directly. \cite{rossi2018closed} models 1-dimensional opinion dynamics on a sphere, \cite{jiang2019degenerate} and \cite{kalimeris2021preference} propose models of behavior shift due to content exposure and content consumption, respectively. Each of these theoretical contributions predicts polarization and extremism of user preferences.
\paragraph{Psychology-Informed Recommendation}
Recent surveys \cite{jesse2021digital,lex2021psychology} cover ways in which recommenders incorporate and account for psychological effects. The work by \citet{lex2021psychology} focuses mainly on ways in which affective, cognitive and personality factors impact user engagement. In contrast to our work, it does not cover the feedback loop between a recommendation systems and user preferences. \cite{jesse2021digital} reviews psychologically-informed recommenders for behavioral priming and digital nudging; it point out a lack of research into psychological foundations of dynamic preference models.

%%%%% EXAMPLES. %%%%%%%%%
\section{Psychological Effects}\label{sec:examples}
This section introduces three psychological effects, Mere-Exposure, Operant Conditioning, and Hedonic Adaptation. For each effect, we will first introduce a classical example of the effect, give general references, and then provide experiments most relevant to recommendation systems. While we find strong connections for the Mere-Exposure and Hedonic Adaptation effects, we will introduce a third effect, Operant Conditioning, whose experimental evaluation allows for a less straightforward connection to recommendation systems. As a running example in this section, we will consider Alice, a user that initially does not like sports content but it is exposed to it by a recommendation system.
\subsection{Mere-Exposure}
Mere-Exposure or the familiarity effect says that humans tend to like more what they are exposed to more often. A classical experiment, cited in \cite{zajonc1968attitudinal}, is the Black Bag Experiment. In 1968 at Oregon State University, C. Goetzinger let a person fully covered with a black bag participate in a course. Other students in the course were hostile at first, but later became friendly towards the person covered with the bag. In a recommendation system context, Mere-Exposure may mean that reactions to content become more favorable after exposure to and/or consumption of similar content. In the case of Alice, whose initial preferences are not favoring sports, a Mere-Exposure effect would predict that with repeated exposure or consumption she becomes more familiar with sports content and starts to appreciate it more.

Mere-Exposure effects are well-established in psychology. \cite{bornstein1989exposure} conducts a meta-study of 200 studies of research in the first three decades following its introduction in \cite{zajonc1968attitudinal}. Much of the research on Mere-Exposure effects is using experimental setups in which subjects are exposed to non-meaningful stimuli, such as Japanese characters for non-Japanese speakers \cite{dechene2009mix}. While this allows to control for prior exposure to the content, such research is not directly meaningful for understanding the Mere-Exposure effect in recommendation systems.

Most relevant to recommendation systems is Mere-Exposure research on advertisement and audiovisual content. An illustrative example of this kind is \cite[Experiment 1]{hekkert2013mere}. In it the experimenter repeatedly shows users images of fictitious, but plausible, products (e.g., a smoke filter) of different aspect ratios. Users' reported rating of attractiveness of aspect ratios of products increased significantly with the number of times it has had been shown to the user. On average each additional exposure led to a 0.2 points increase on a 7-point scale. (The experiments in \cite{nordhielm2002influence, fang2007examination} for the exposure to banner ads and \cite{cox2002beyond} for exposure to pictures of fashion designs made similar findings.) The study found as well that \enquote{conspicuous} exposure, i.e. exposure that draws attention to the unfamiliarity of the product, leads to a smaller Mere-Exposure effect (0.03 points increase per additional exposure).

\subsection{Operant Conditioning}
Operant Conditioning is the effect that beings tend to engage more in activities that are associated with positive stimuli (positive reinforcement) and avoid activities associated with negative stimuli (negative reinforcement). In a classical experiment from \cite{skinner2019behavior}, B.F. Skinner puts a hungry rat into box containing a food dispenser and a button. Food is released whenever the button is pressed. If the rat does not press the button, it gets a small electric shock. As an observation of this experiment, the rat presses the button more and more frequently. In recommendation systems, Operant Conditioning predicts higher engagement with content that was \enquote{surprisingly good}  and less that is associated with content that was \enquote{surprisingly bad}. If the initially sports-averse Alice reacts according to Operant Conditioning, depending on her baseline level, she might be underwhelmed by the sports content, and dislikes it more after being exposed to it, or her baseline is very low, in which case she might start liking it.

Operant Conditioning is well-documented in Behavioral Psychology, both in animals and humans. Skinner \cite{skinner2019behavior,ferster1957schedules} conducted several studies in animals which finds evidence for Operant Conditioning. To our knowledge the first study of Operant Conditioning in humans is  \cite{fuller1949operant} which conditioned a young man with a developmental disorder to raise his arm. We refer the reader to the monograph \cite[Chapter 11-13]{cooper2007applied} for a treatment of Operant Conditioning and the following behavioral movement in psychology.

While much of the work on Operant Conditioning is conducted in animal experiments, work that most closely resembled a recommendation systems is in consumer psychology. \cite{foxall2004consumer} (see surveys \cite{foxall2010invitation,foxall2017behavioral} for follow-up work) introduced the Behavioral Perspectives Model, which classifies consumer choices into several \enquote{reinforcers} which might depend on the product-dependent, social, or monetary (shopping online is often costly) consequences of making consumption decisions. As an exemplary experiment, \cite{fagerstrom2011study} considers the effect of externally provided reinforcers (e.g. shipping cost, shipping duration, or price) on consumer choice among two different online shops. \cite{fagerstrom2011study} report that a significant fraction of the subjects followed the (positive) reinforcements set by the experimenter. 
We note that an interpretation of Operant Conditioning for dynamic \emph{preference} models is challenging. On the one hand, behavioralism views behavior purely as a black box, which is why the effect is framed around behavior, not preferences. Our translation to preferences requires assuming that Operant Conditioning may lead to changes in preferences. Second, the definition of baseline we will consider in the quantitative model below depends on past preferences, which is closer to literature on adaptation, e.g. the quantitative model of \cite{helson1947adaptation}, than of Operant Conditioning.

\subsection{Hedonic Adaptation}
Hedonic Adaptation is the effect that after some time, any change in happiness fades, and humans return to a baseline level. In a classical study \cite{brickman1971hedonic}, P. Brickman asked lottery winners and paraplegics for their happiness and how they expect their happiness level to be in a year, finding that major life events had negligible effect on their happiness. Hedonic Adaptation means that engagement with content returns to a baseline level after some time. If Alice's \enquote{baseline self} does not like sports content, she will return to this baseline irrespective of the content recommended---which she also would if her \enquote{baseline} self likes sports.

\cite[ch. 16]{kahneman1999objective} reviews many earlier findings on the human adaptation to repeated exposure to noise (inconclusive evidence) as well as incarceration and increased income (supporting evidence). 

In Hedonic Adaptation literature most relevant to recommendation systems, studies consumption scenarios. \cite{chugani2020all}'s experiment lets students choose a sticker and attach it to an everyday object. Eliciting self-reported happiness with the sticker, the study finds a 4.5 point decline on a 100-point scale of reported happiness with a sticker in a 3-day interval. \cite{yang2015sentimental} finds a loss between 1.75 point (\enquote{low sentimental value}) to 7.75 point (\enquote{high sentimental value}) decrease on a 100-point scale for Google Image search result shown to users for 6 short intervals of 10 seconds. \cite{nelson2008interrupted} plays songs and asks for reports of happiness with the song at different points. Still on a 100-point scale, the preference is reduced by 15 points from 10 seconds into the song to one minute into the song. Finally \cite{chugani2015happily} showed paintings to subjects in three 15-second exposure intervals. The exposure led to a reduction in happiness with the painting of about 12 points on a 100-point scale. 

All of these studies have relatively short exposure times. However, content types are comparable to many contemporary recommendation systems, demonstrated effects on self-reported happiness are quite substantial.

% In the next section we attempt to formalize the three effects proposed here.

%%%%% SETUP. %%%%%%%%%
\section{Formalizations of Psychological Effects}\label{sec:setup}
In this section, we propose mathematical formalizations of Mere-Exposure, Operant Conditioning, and Hedonic Adaptation. We start with our basic notation.

At each round, a recommendation system recommends one of $N$  pieces of content, each with an associated $d$-dimensional item vector. Throughout, we assume that the item vectors are fixed.\footnote{This assumption is restrictive, but is justified in content-based recommenders, e.g., when content is featurized in terms of topic, genre, length, political inclination, etc., or high-dimensional content embeddings of images, videos or text based on supervised or unsupervised learning. The assumption that item vectors are fixed also holds approximately in Collaborative Filtering settings if item representations are updated less frequently than user representations.} Since the item representations are known and fixed, it is without loss to consider a single user at a time. Denote by $\pi_t \in \R^d$ the preference vector of the user at time $t$. The user reacts to a piece of content with item vector $v\in \R^d$ according to a rating function, which we assume to be linear, as is common in collaborative filtering, $r(\pi_t,v) \coloneqq \langle \pi_t, v \rangle + \varepsilon_t \in \R$ for independent noise $\varepsilon_t$, on which we will make additional distributional assumptions below. At each time step $t$, the recommender chooses a piece of content with associated latent representation $v_t$.
% as a function of item \emph{scores}. A score is a prediction of an item rating $s( u_t, v)$, where $ u_t \in \R^d$ is the preference vector estimated from past interactions. Given the scores, the recommender select an item according to some (potentially stochastic) function of the scores. We will consider score estimates that are linear in estimated preferences, $s(u_t, v) = \langle u_t, v \rangle$.

In contrast to static recommendation system models, we model user preferences change due to content exposure and consumption: $\pi_{t+1} - \pi_t = f(\hist_t)$ for some user-item history $\hist_t = (\pi_1, v_1, \pi_2, v_2, \dots, \pi_t, v_t)$, and a potentially random function $f$. For a static user, $f(\hist_t)$ is constant $0$.\footnote{Note that our model does not differentiate between the effects of exposure to and consumption of content. A more general model would model consumption probabilities, and allow for separate effect strengths of exposure and consumption.}

Next, we propose quantitative models for the psychological effects introduced in Section \ref{sec:examples}.

\begin{figure}
    \centering
    \begin{subfigure}[b]{.23\textwidth}
    \centering
        \includegraphics[height=2.5cm]{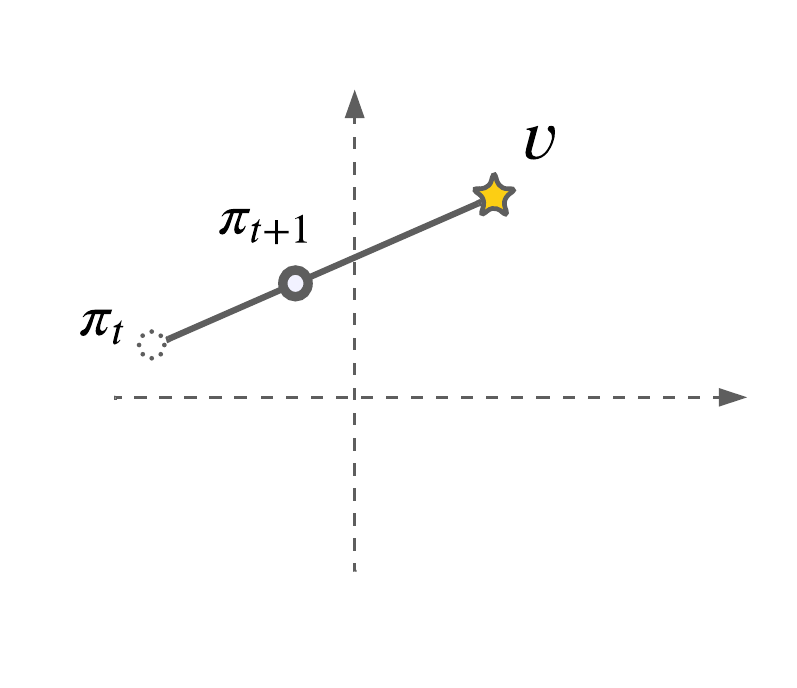}
        \caption{Mere-Exposure}\label{subfig:me_depiction}
    \end{subfigure}
    \begin{subfigure}[b]{.23\textwidth}
    \centering
        \includegraphics[height=2.5cm]{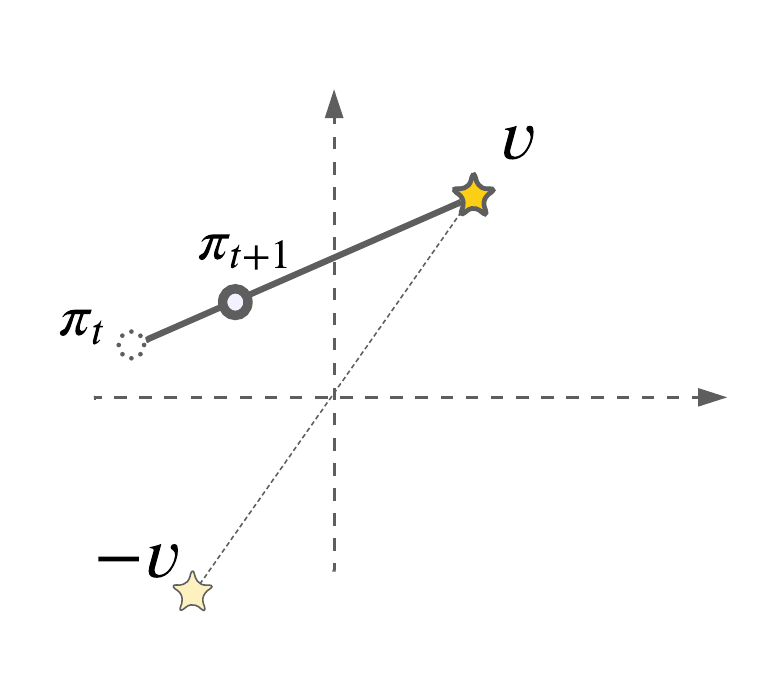}
        \caption{pos. Operant Conditioning}\label{subfig:posoc_depiction}
    \end{subfigure}
    \begin{subfigure}[b]{.23\textwidth}
    \centering
        \includegraphics[height=2.5cm]{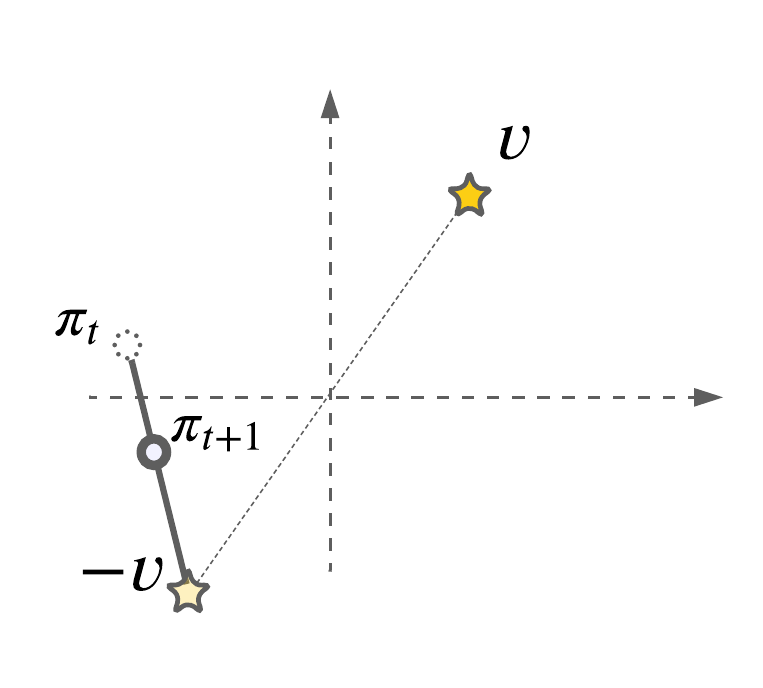}
        \caption{neg. Operant Conditioning}\label{subfig:negoc_depiction}
    \end{subfigure}
    \begin{subfigure}[b]{.23\textwidth}
    \centering
        \includegraphics[height=2.5cm]{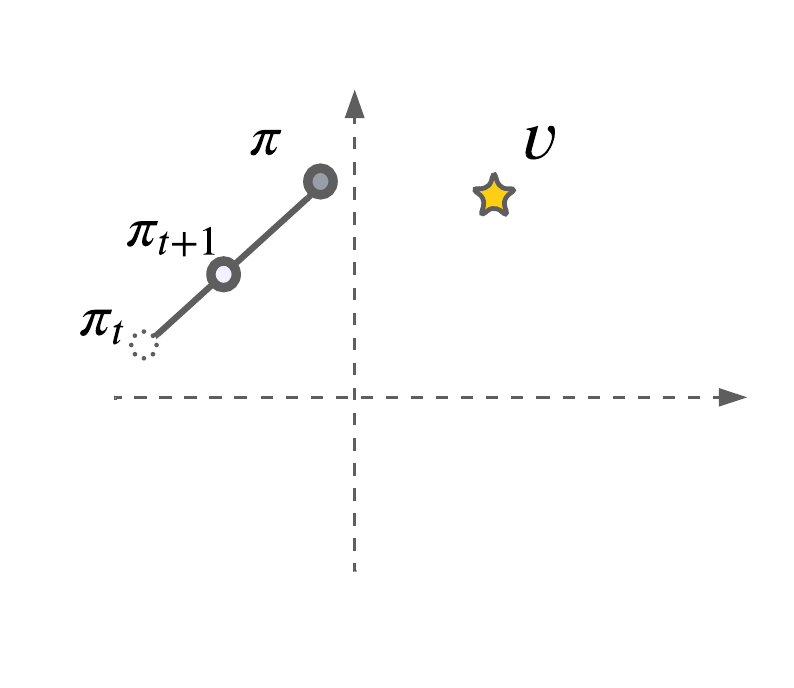}
        \caption{Hedonic Adaptation}\label{subfig:ha_depiction}
    \end{subfigure}
    \caption{Updates in preference space for Mere-Exposure, Operant Conditioning and Hedonic Adaptation. Mere-Exposure moves preference vectors a constant fraction of the distance towards the recommended content; Operant Conditioning either moves towards, or away, depending on the direction and magnitude of \emph{surprise}. Hedonic Adaptation leads to convergence towards a baseline rating. %We denote by $\pi_0, \pi_{t}$ and $\pi_{t+1}$ the initial preference vector, preference vector at time $t$ and $t+1$ respectively; $v$ is the vector embedding corresponding to the recommended item.
    }
    \label{fig:depictions}
\end{figure}
\subsection{Mere-Exposure}\label{subsec:mepred}
For Mere-Exposure, we consider the linear model 
\begin{equation}
    \pi_{t+1} - \pi_t =\gamma f_{\ME} (v_t, \pi_t) = \gamma (v_t - \pi_t)
\end{equation}
for some $\gamma \in [0,1]$, compare \autoref{subfig:me_depiction}. Whenever an element $v_t$ is shown, the preference vector moves a $\gamma$-fraction of the line between $u_t$ and $v_t$. This captures the idea that whenever users are exposed to content, this makes them like this content more. %We note that this model is related to the one studied in \cite{passino2021next}, which does not make its connection to psychological theory explicit and does not consider behavior in simulations, which are the main contributions of this paper.

Our model is parameterised by a single quantity that determines how much a user moves into a certain direction. The strength of preference movement might depend, given the psychology literature reviewed above, on the conspicuousness of content exposure.

\subsection{Operant Conditioning}
The space of possible models that capture Operant Conditioning is large, and we consider a particular parameterization, highlighting some of the qualitative features of positive and negative reinforcement. The model we consider here captures that users will adjust their reaction to content that surprises them:
\begin{align*}
\pi_{t+1} - \pi_t &= \gamma f_{\OC} (\hist_t) \\
&= \gamma \lvert\surprise(\hist_t)\rvert ( \sgn (\surprise (\hist_t)) v_t -  \pi_t)
\end{align*}
for $\gamma \in [0,1]$, compare \autoref{subfig:posoc_depiction} for the case of positive reinforcement, $\sgn(\surprise(\hist_t)) = 1$, and \autoref{subfig:negoc_depiction} for the case of negative reinforcement, $\sgn(\surprise(\hist_t)) = -1$. The magnitude of the preference shift is scaled by the size of a \emph{surprise} term, and the direction of the change is determined by the sign of the surprise. If the surprise is positive then the preference moves in the direction of the item $v$. Conversely, if the surprise is negative then the preference moves towards $-v$.

The surprise is a function of difference between a baseline level of engagement and the current rating. We model the expected engagement as an discounted average of historical ratings, and use an $\arctan$ function to map to the range $[-1,1]$. This yields the surprise term of the form:
\[
\surprise(\hist_t) :=\arctan \left(\frac{\sum_{\tau=1}^{t-1} \delta^\tau r_{t-\tau}}{\sum_{\tau=1}^{t-1} \delta^\tau} - r_t \right), \delta \in [0,1].
\]
The choice of an exponential decay is motivated in psychology and neuroscience, compare, e.g., habituation \cite{marsland2009using}.

\subsection{Hedonic Adaptation}
For hedonic adaptation, we propose a fairly simple model: a linear drift towards a constant \emph{baseline preference vector} $\pi$:
\[
\pi_{t+1} - \pi_t = \gamma f_{\HA}(\pi_t) = \gamma (\pi - \pi_t)
\]
for $\gamma \in [0,1]$, compare \autoref{subfig:ha_depiction}. The update moves the user towards a (fixed) baseline preference $\pi \in \R^d$. Note that this behavioral shift is irrespective of the recommended content $v_t$. 

A shared property of all proposed models is that they do not lead to arbitrarily large preference vectors.
\begin{prop}\label{prop:bounded}
Let $f \in \conv(f_{OC}, f_{HA}, f_{ME}, 0)$; i.e $\pi_{t+1} - \pi_t = \gamma_{ME} f_{ME}(\pi_t, v_t) + \gamma_{OC} f_{OC}(\pi_t, v_t)+ \gamma_{HA} f_{HA}(\pi_t)$ where $\gamma_{ME} + \gamma_{OC} + \gamma_{HA} \le 1$. Then for any sequence of recommendations, $\pi_t$ is bounded (see proof in \autoref{sec:proofs}).
\end{prop}
%The proof of this statement is in the supplementary materials.

%%%%%% QUALITATIVE PREDICTIONS. %%%%%%
\section{Experiments}\label{sec:predictions}
This section presents simulation results for the formalizations $f_{\OC}, f_{\ME}$, and $f_{\HA}$ of Operant Conditioning, Mere-Exposure, and Hedonic Adaptation. We use our results to compare against  psychological evidence.
\subsection{Experimental Setup}
\paragraph{User behavior}
We sample $N$ i.i.d item vectors from a multivariate normal distribution, $v_i \sim \mathcal{N}(0, \sigma I_{d})$. The initial user preference $\pi_0$ is sampled from the same distribution.

At time $t$, the recommender selects an item $v_t$ based on an estimate of current ratings $s_{it} = \langle u_t, v_i\rangle, i=1, 2, \dots, N$.
The user observes the recommended item and responds with a rating $r(\pi_t, v_{t}) = \langle \pi_t, v_t\rangle + \epsilon$. As a result of exposure to and/or consumption of the content the user preference updates to $\pi_{t+1} = \pi_t + \gamma f(\pi_t, v_{t}) + \epsilon_{t}'$, where $\epsilon'_t \sim \mathcal{N}(0, \sigma' I_d)$ is zero mean stochastic noise applied to the preference dynamic. We add noise to avoid unstable equilibria states.
\paragraph{Preference Estimation}
We make recommendations based on estimates of user preference: $u_t$. We initialize the estimate with a random multivariate normal vector: $u_0\sim \mathcal{N}(0, \sigma I_d)$. Given a recommended item $i$, and observed rating $r_{it}$ we update the representation of the user preference estimate $u_t$ according to Online Gradient Descent (OGD) \cite{hazan2016introduction} for the loss function $\ell_t(u) = \frac{1}{2}\left((r_t - \langle u, v_i\rangle)^2 + \eta \Vert u\Vert^2\right)$
\[u_{t+1} = u_t - \alpha \nabla_{u_t} \ell_t(u_t) = (1-\alpha \eta)u_t + \alpha (r_{it} - s_{it}) v_{i}.\]

In \autoref{app:additional} we repeat our experimental setup in the \emph{oracle} model in which the recommender has direct access to preferences. We find that the qualitative insights from this section hold both in the oracle and in the estimation model; thus the dynamic patterns that we observe are primarily driven by behavioral shift rather than by estimation error.
\paragraph{Item Selection}
We consider three baseline recommenders and softmax selection rule with with different temperatures:
\begin{enumerate}
    \item Baselines: The \uniform{} selection chooses an item uniformly at random from the set of items. The 
    \constant{} selection repeatedly selects the same item for all recommendation rounds. \greedy{} selection picks the item with the maximum predicted score $i_t^*:=\arg\max_i s_{ti}$, breaking ties randomly.
    \item \softmax{} selection: Given predicted scores $\{s_{it}\}_{i=1}^N$ and the estimate of the preference vector $u_t$, a \softmax{} selection rule with temperature $\beta$ selects item $i$ with probability\footnote{We scale the $\beta$ parameter by the norm of the estimated preference to maintain the expected engagement between consecutive rounds constant.}
    \[\mathbb{P}[i_t^* = i] \propto \exp\left(\frac{\beta }{\Vert u_t\Vert} s_{it}\right).\]
\end{enumerate}
In each of the following illustrations of preference trajectories, we will depict the long-term preference distribution using a cloud, while the first moves of preferences are depicted by connected dots.
\subsection{Qualitative Behaviors and Testable Predictions of Proposed Behavioral Models}
\subsubsection{Mere-Exposure}
\begin{figure*}
    \centering
    \begin{subfigure}[b]{0.9\textwidth}
         \centering
         \includegraphics[width=\textwidth]{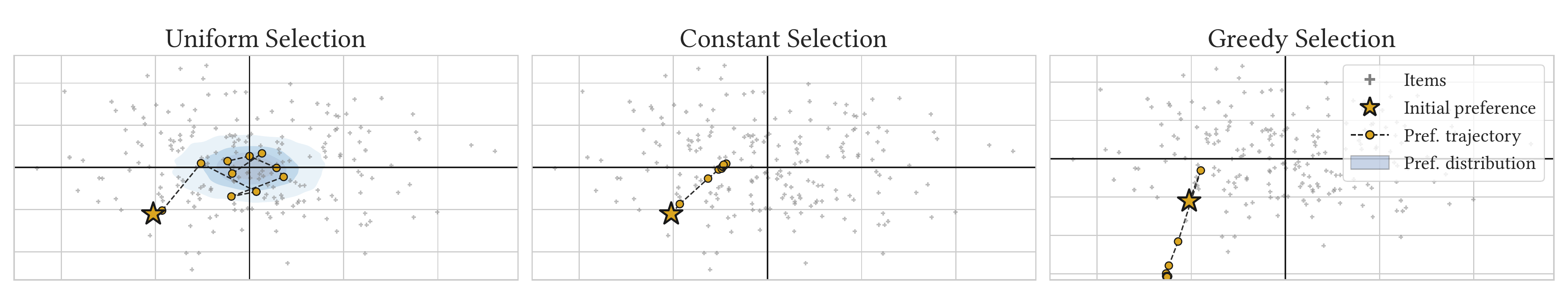}
         \caption{\textbf{Baseline recommenders}: \uniform{}, \constant{} and \greedy{} selection rules}
         \label{fig:me_baseline}
     \end{subfigure}
     \begin{subfigure}[b]{1\textwidth}
         \centering
         \includegraphics[width=0.9\textwidth]{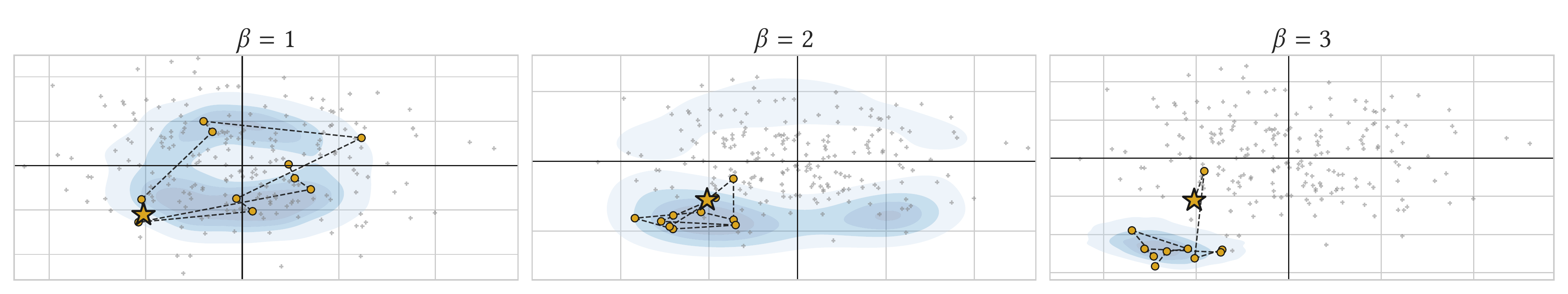}
        \caption{\textbf{\softmax{} recommenders}:  $\beta = 1, 2, 3$}
         \label{fig:me_softmax}
     \end{subfigure}
    \caption{\textbf{Preference Trajectory ($d=2$, $N=1000$)}: Mere-Exposure dynamic $\gamma_{ME} = 0.1$. Note preferences stay stochastic and lead to a long-term distribution around the origin for the \uniform{}. \constant{} and \greedy{} converge to a repeatedly recommended item.}
    \label{fig:me}
\end{figure*}

Under this dynamic, users move in the direction of the recommended content, irrespective of how highly they rate it. \autoref{fig:me} displays user trajectories in a 2-dimensional preference space for $f_{\ME}$ with $\gamma_{ME} = 0.1$. In the case of \uniform{} recommendations, the preferences converge to a ball centered at the origin. For the \greedy{} and \constant{} baseline recommenders, the user preference converges to the latent representation of the item that is repeatedly recommended. With the \softmax{} selection rule, we observe that the long-term distribution of preferences over time traverses the item space. Under selection rules that favor exploration, for instance \softmax($\beta = 1$), the preference vector moves faster around the item space, yet stays closer to the origin, compared to selection policies that favor expected engagement; e.g. \softmax($\beta =3$).

\begin{figure}
    \centering
        \begin{subfigure}[t]{0.42\textwidth}
        \vskip 0pt
        \centering
        \includegraphics[height =3.2cm]{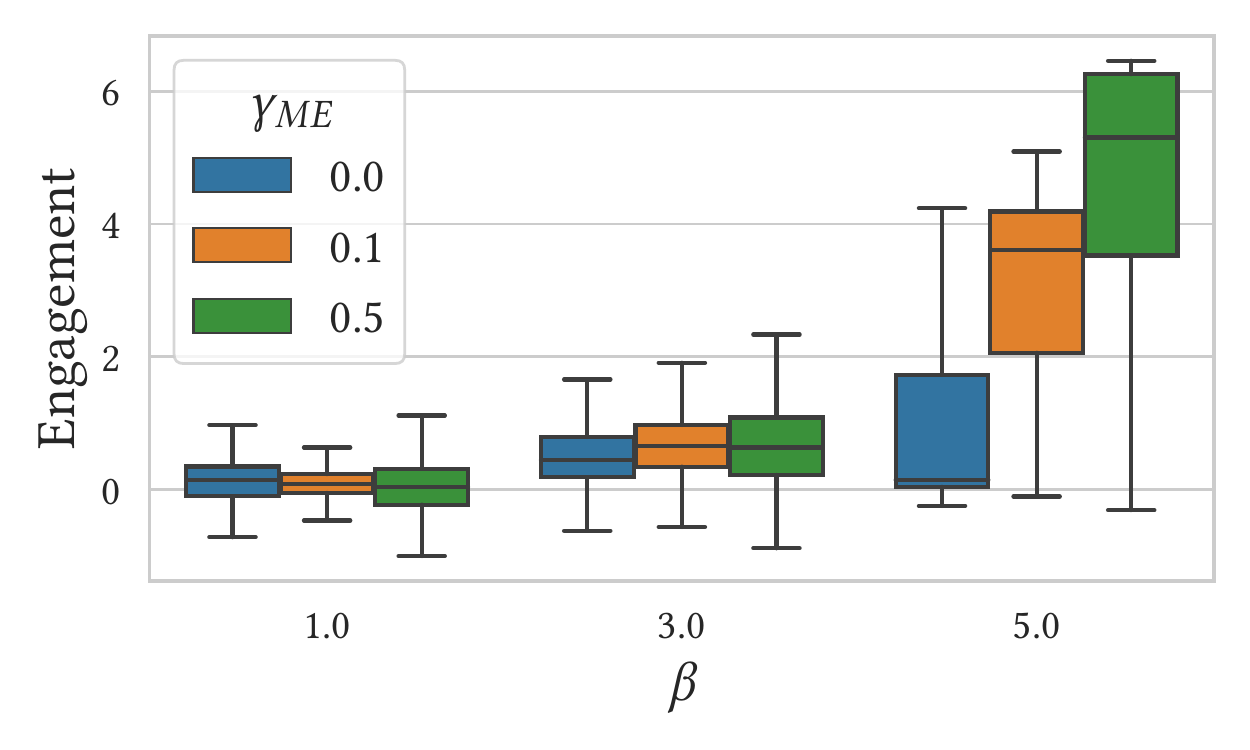}
        \caption{Large $\gamma_{ME}$ leads to higher engagement for high $\beta$.}
    \end{subfigure}
    \begin{subfigure}[t]{0.20\textwidth}
        \vskip 0pt
        \includegraphics[height=3.2cm]{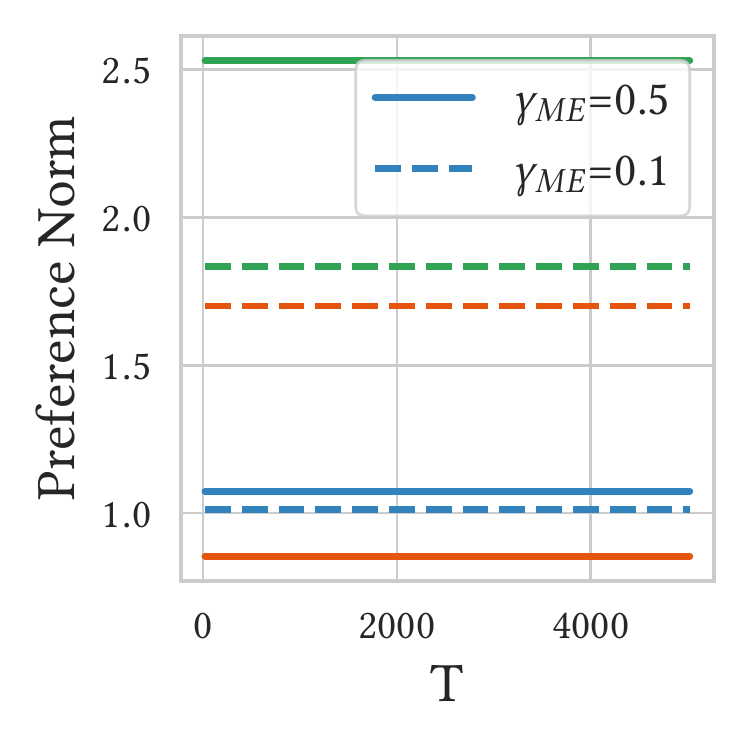}
        \caption{Magnitude over time.}
    \end{subfigure}
    \begin{subfigure}[t]{0.27\textwidth}
        \vskip 0pt
        \includegraphics[height=3.2cm]{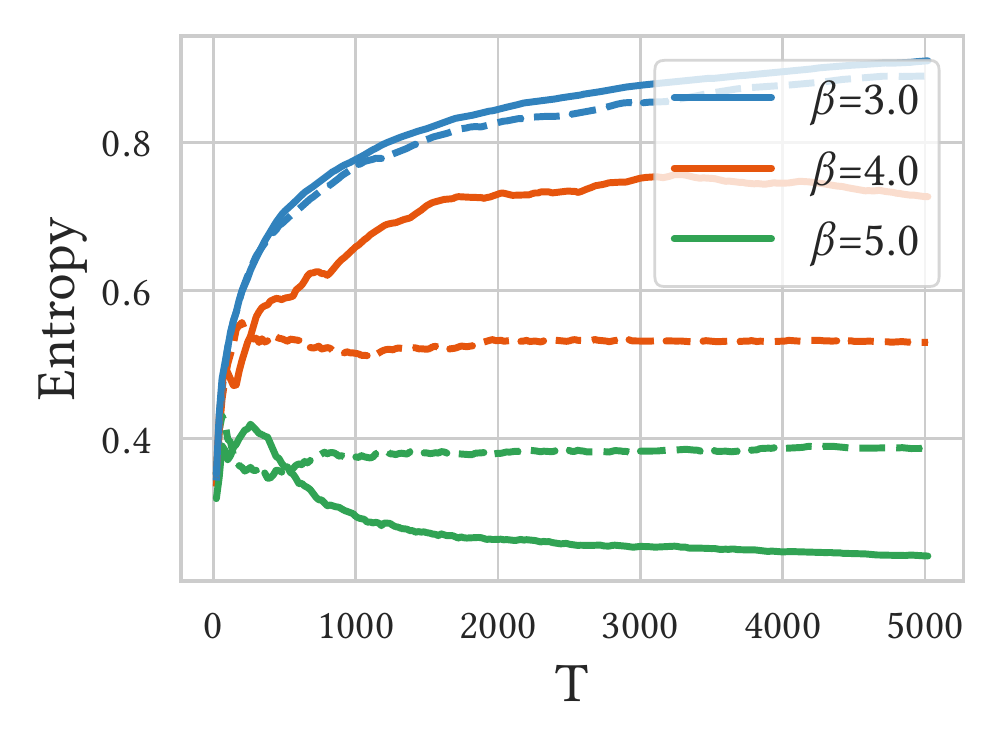}
        \caption{Entropy over time. }
        \label{subfig:entropy}
    \end{subfigure}
    \caption{\textbf{Higher Dimensions ($d=8$, $N=5000$):} Dependence of user engagement, preference magnitude, and diversity of consumption for different \softmax{} $\beta$ parameters. (a) A higher-$\beta$ softmax leads to higher engagement, which is exaggerated by stronger Mere-Exposure. (b) A very high-$\beta$ \softmax{} ($5.0$) leads to preferences of high norm. (c) The entropy of consumed content may be high even for moderately high-$\beta$ \softmax{} ($4.0$).}
    \label{fig:me_highd}
\end{figure}

\autoref{fig:me_highd} shows how the $\beta$ parameter affects the engagement of the user, the magnitude of their preference and the the diversity of their content consumption, operationalized as entropy of the distribution of recommended content. We first observe that engagement may increase due to Mere-Exposure (see the increase of engagement for constant $\beta$ and increasing $\gamma_{\ME}$). We further note that a recommender with $\beta=5$ has a very high engagement in particular with high Mere-Exposure. This might raise the question of whether engagement is a valid metric for users with dynamic preference. We will consider this question further in \autoref{sec:design}.

Here we derive the testable prediction that for \softmax{} selection rules and Mere-Exposure dynamics the estimates of the preference vector stay relatively constant over time, and the magnitude of the preference increases with the temperature parameter $\beta$.

\subsubsection{Operant Conditioning}
When the preference shift is governed by Operant Conditioning, we observe that the norm of the preference vector oscillates. This phenomenon can be explained by the type of reinforcement that the Operant Conditioning $f_{\OC}$ induces. It is illustrative to analyze the behavior for the \greedy{} recommender. In the beginning,  the user is served a recommendation for which she will respond positively. Indeed the surprise term is positive since the expected engagement is 0, and thus the user preference will shift in the direction of the item. With the preference moving towards the item, its score increases and the \greedy{} selection picks it again. The positive reinforcement of preferences from the previous round  ensures that the surprise term is still positive; leading to further amplifications in the preference of the  direction of the item. Eventually, given the increases in the expected engagement from the previous round, the expected surprise goes to 0. At this time any noise in the response can make the surprise term negative and thus sending the preference vector in the direction of $-v$. As $-v$ is nearly diametrically opposed to the preference at this time, even a small negative surprise would considerably decrease the magnitude of the preference vector. However, since the movement is confined to the direction of the original preference vector, the \greedy{} selection will keep recommending it. As the preference decreases in magnitude, so does the engagement of the user. However, as expected engagement is a lagging metric, the surprise term becomes even more negative, thereby creating a downwards spiral which ends with the user getting completely bored and losing their preference $(\pi_t \approx 0$). After enough time-steps the historical expectation decreases enough such that some other direction (by random chance) has a positive surprise term, commencing again the amplification of the preference in that direction.\footnote{It might stabilize at high values of engagement if there is no noise in the preference updates.}

\begin{figure*}
    \centering
    \includegraphics[width=0.9\textwidth]{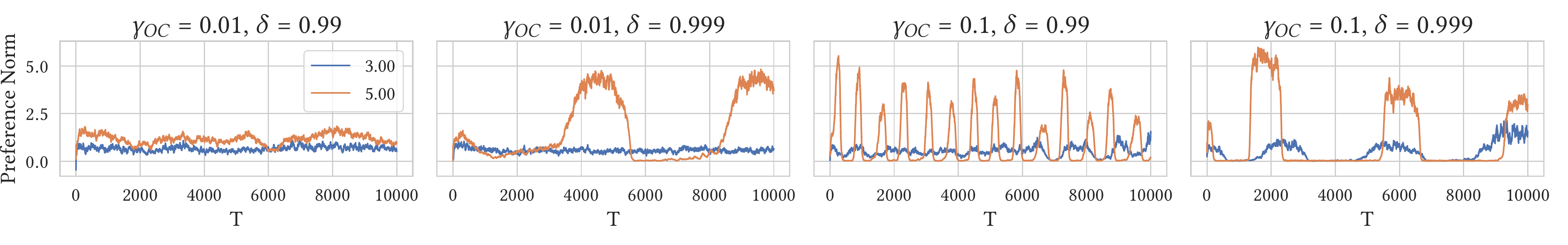}
    \caption{\textbf{Magnitude of Scores in Higher Dimensions ($d=8$, $N=5000$)}: Across behavioral models, higher-engagement selection policies (high $\beta$) correspond to more extreme oscillations. When $\OC$ effects are large ($\gamma_{OC}=0.1$), the discount factor $\delta$ has a significant impact on the period of oscillations. Lower $\delta$ corresponds to a recency bias, where older ratings play a diminished role in forming baseline expectations for engagement, and thus lead to oscillatory patterns of higher frequency.}
    \label{fig:oc_period}
\end{figure*}

The \softmax{} selection rules are less extreme version of the \greedy{} recommender, yet many of them still show oscillatory patterns. The period and amplitude of these oscillation depends on the \softmax{} parameter $\beta$ and on the decay parameter in the Operant Conditioning update, $\delta$. The larger the $\beta$, the larger is the amplitude of the preference swings and the shorter is the period, compare \autoref{fig:oc_period}.

The oscillations seen in the simulations are testable predictions of Operant Conditioning (note that \autoref{fig:oc_period} shows estimated scores, not unobserved user preferences). The review of psychological literature in  \autoref{sec:examples} did not show evidence of such oscillatory patterns in consumption under Operant Conditioning, which makes testing this model with data on a deployed recommender particularly important.

\begin{figure*}
    \centering
    \begin{subfigure}[b]{1\textwidth}
         \centering
         \includegraphics[width=0.9\textwidth]{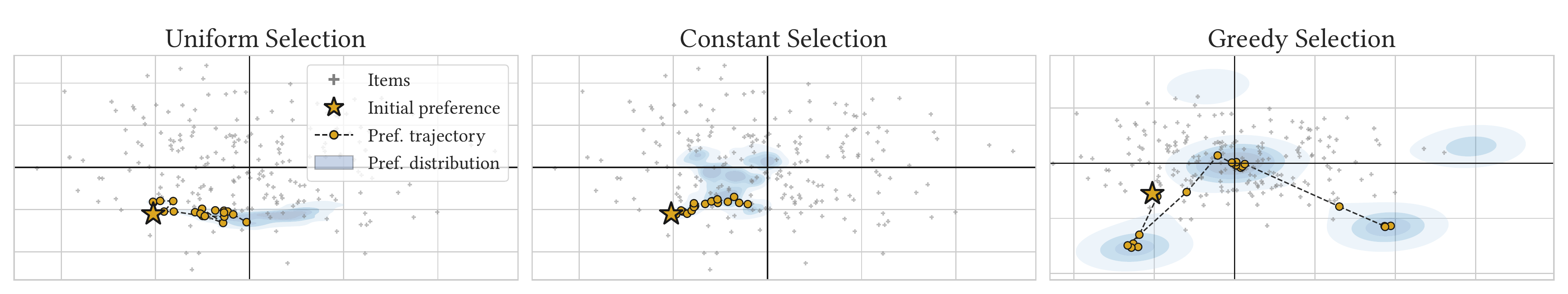}
         \caption{\textbf{Baseline Recommenders}: \uniform{}, \constant{} and \greedy{} selection rules}
         \label{fig:oc_baseline}
     \end{subfigure}
     \begin{subfigure}[b]{1\textwidth}
         \centering
         \includegraphics[width=0.9\textwidth]{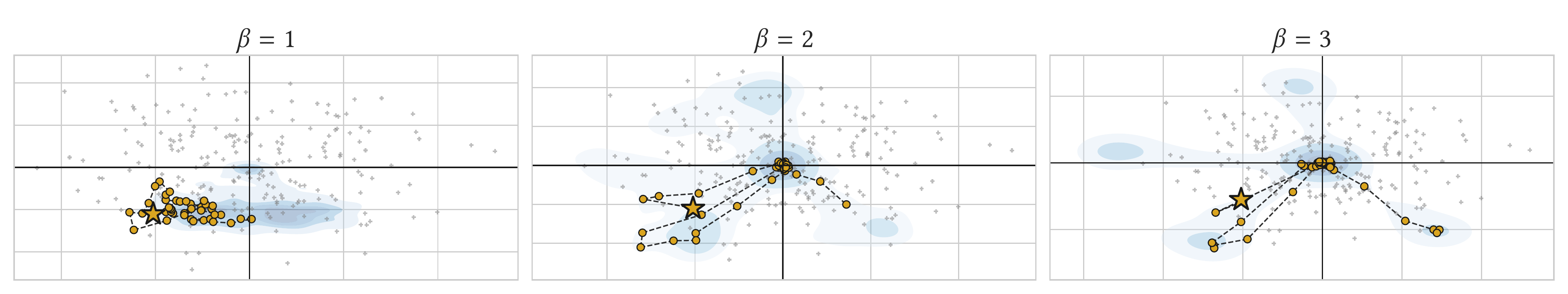}
        \caption{\textbf{\softmax{} Recommenders}:  $\beta = 1, 2, 3$}
         \label{fig:oc_softmax}
     \end{subfigure}
    \caption{\textbf{Preference Trajectory ($d=2$, $N=1000$)}: Operant Conditioning, $\gamma_{OC} = 0.1$. Preferences oscillate for \constant{} and \greedy{} selection.}
    \label{fig:oc}
\end{figure*}

\subsubsection{Hedonic Adaptation}
\begin{figure*}
    \centering
    \begin{subfigure}[b]{1\textwidth}
         \centering
         \includegraphics[width=0.9\textwidth]{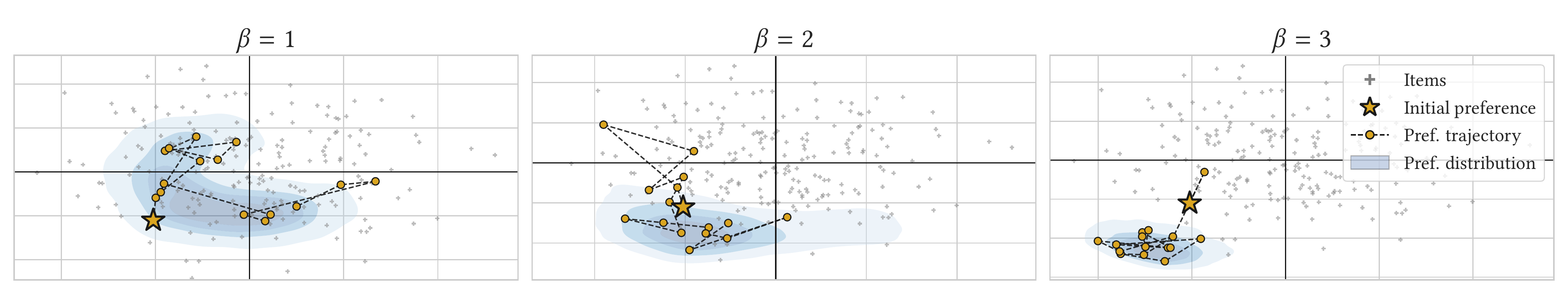}
         \caption{Mere-Exposure with Hedonic Adaptation: \softmax{} recommenders}
         \label{fig:me_ha}
     \end{subfigure}
     \begin{subfigure}[b]{1\textwidth}
         \centering
         \includegraphics[width=0.9\textwidth]{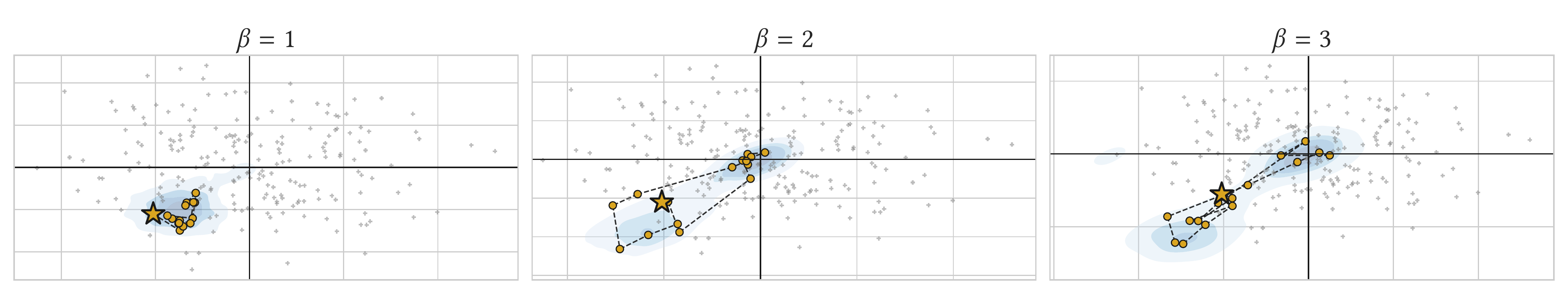}
         \caption{Operant Conditioning with Hedonic Adaptation; \softmax{} recommenders}
         \label{fig:oc_ha}
     \end{subfigure}
    \caption{\textbf{Preference Trajectory ($d=2$, $N=1000$):} Hedonic Adaptation $\gamma_{HA} = 0.01$. Hedonic adaptation biases the trajectories and long-term preference distribution in the direction of the long-term preference distribution.}
    \label{fig:ha}
\end{figure*}

Pure Hedonic Adaptation leads to convergence to the baseline point, and does so independently of the recommendation policy. In combination with other effects, Hedonic Adaptation limits the amount by which other effects are perceived. For example when combined with Mere-Exposure (\autoref{fig:me_ha}) Hedonic Adaptation provides a strong drift towards the baseline preference; and thus the user preference travels \enquote{less} within item space. In joint dynamic with Operant Conditioning (\autoref{fig:oc_ha}), user preferences still oscillate, but the oscillations are limited to the direction of the baseline preference.
In both cases, Hedonic Adaptation biases, but does not overwhelm the dynamics observed for Mere-Exposure (traveling through preference space) and Operant Conditioning (oscillations). Biases towards part of the item space are testable predictions of $f_{\HA}$ in combination with other behavioral models.

\subsection{Using Testable Hypotheses to Critique Dynamic Preference Models}\label{subsec:refinining}
The testable predictions of Mere-Exposure and Operant Conditioning are quite distinct when interacting with recommenders. $\ME$ predicts that user ratings will be fairly constant even if recommended content changes over time; whereas $\OC$ predicts oscillatory patterns in the ratings. These predictions may, or may not, be in line with findings in psychology or observed in deployed recommendation systems. Both types of checks of models allow for further refinement of user behavioral models. 

First, the qualitative behaviors found may be challenged, potentially motivating other quantitative models. The oscillation pattern we observed for our quantitative model of Operant Conditioning, $f_{\OC}$ is, to our knowledge, not known in psychology. This might rely on the fact that $\softmax{}$ recommendation is a novel type of repeated stimulus unknown in psychology, or might point to a weakness of the model we proposed. Validating such models on deployed system can help select or reject formalizations of models.

Further, refinements of functional forms are also possible. Under $f_{\OC}$, negative reinforcement is stronger than positive reinforcement, as our experiments in this section showed. User preferences in recommendation systems might not exhibit this asymmetry, and motivate new models for Operant Conditioning, e.g., decreasing the slope of the surprise when surprise is negative. Similarly, the fact that the preference lingers around 0 for several time periods suggests that our formulation of users' surprise over-emphasizes early exposure. This suggests a further refinement on our model by reducing the discount factor $\delta$.

\section{Discussion and Implications}\label{sec:discussions}
This article proposes psychologically grounded models and derives their implications. We first discuss our approach to grounding models, and place them into a bigger methodology. We then further investigate our observation that user models might affect recommendation metrics in design.
%%%%%% METHODOLOGY.  %%%%%%%%%
\subsection{A Methodology for Psychologically Grounded Dynamic Preference Models}\label{sec:plausibility}
The approach taken in the current article followed several steps, compare \autoref{fig:vignettes}.
\begin{description}
    \item[Statement] Declare a psychological effect and review relevant psychology literature (\autoref{sec:examples});
    \item[Formalization] Formalize the effect within a recommendation system model (\autoref{sec:setup});
    \item[Predictions] Inspect properties of the model using theory, simulations, or a combination of the two to derive testable predictions (\autoref{sec:predictions});
    \item[Data] Test the predictions of the models against historical and/or interventional data in a deployed recommendation system (not performed for this study);
\end{description}
Predictions and data may be used to refine the formalization chosen for a particular effect (\autoref{subsec:refinining}).

In addition to the application of the proposed methodology to modelling concrete effects, it may be used to reconsider some models proposed in the recent literature on dynamic preference models in recommendation systems. Next we give three examples of such a discussion.

\begin{example}
In \cite{passino2021next} the authors consider and evaluate several dynamic user models. The authors write that \enquote{an exponentially weighted moving average [\dots] distribution is obtained within
each time window}. The paper considers an update $u_{t+1} = (1-\gamma)u_t  + \gamma v_t$, where the item vectors $v$ are modeled as unit vectors with respect to music genre, and the user factors is a probability vector encoding the likelihood that a user will consume a content from that genre. While this model is functionally equivalent to Mere-Exposure, \cite{passino2021next} presents the model in a purely mathematical description. Our review on Mere-Exposure might allow, for example, to make a model context-adaptive: If users are detected as listening in the background (conspicuous Mere-Exposure), $\gamma_{\ME}$ is increased.
\end{example}
\begin{example}\label{ex:engagement}
The authors of \cite{carroll2021estimating} assume a user model based on a \emph{theory of chosen preferences} \cite{bernheim2021theory}. The authors of \cite{carroll2021estimating} describe \cite{bernheim2021theory}'s (metacognitive) model as: \enquote{on a high-level, at each timestep users choose their next-timestep preferences to be more \enquote{convenient} ones---ones which users expect to lead them to higher engagement value.} While \cite{bernheim2021theory} gives examples of how their models explain several behavioral effects (conformism, closed-mindedness, sour grapes---the psychological effect that things that are unattainable are less liked), among others, \cite{carroll2021estimating} does not discuss whether this effect, and the particular cognitive model, is relevant in a recommendation system. Explicit psychological models would have allowed to identify parameter ranges for which users following \cite{bernheim2021theory}'s theory resemble Mere Exposure, or another psychological effect.
\end{example}
\begin{example}
Having structural models allows critiquing the precise formulation of a psychological effect. \cite{jiang2019degenerate, kalimeris2021preference} study preference dynamics in closed loop feedback with recommendations and argue theoretically and via simulations that recommendation systems lead to amplification of preferences and consequently to radicalization of users. In \cite{jiang2019degenerate} the authors model the effects that repeated recommendation of an item have on preferences. They propose a model which bears similarity to our Mere-Exposure model and concludes that recommendations lead to unbounded preferences. As unbounded interest in a certain type of content is not a plausible prediction, one can critique the proposed user drift dynamics.  \cite{kalimeris2021preference} proposes a model akin Operant Conditioning and argue for the extremization of user preferences by proving divergence of the preference estimates. The structural model of preference update is formalized in such a way that the estimated preference vectors impact the true preferences directly, which might not capture the correct causal relationship between recommendations and user preferences.
\end{example}
%%%%%% DESIGN.  %%%%%%
\subsection{Evaluation and Design for Dynamic Users}\label{sec:design}
In this section we consider how dynamic preference models may affect recommendation system evaluation metrics and design. This is based on the observation in \autoref{sec:predictions} that \softmax{} recommenders were able to attain high levels of both engagement and diversity. In dynamic settings we illustrate with an example that a recommendation algorithm that improves both engagement and diversity might have unintended consequences.

We will measure engagement as the average rating over time and diversity as the entropy of the normalized counts of the consumed items. When user have static preferences the \softmax{} recommender is known to make the optimal trade-off between the expected ratings and the entropy of item selection probabilities, compare, e.g., \cite{jaynes1957information}.

\begin{theorem}\label{thm:boltzmann}
For any finite set of items $v_i, i=1, 2, \dots, N$ with ratings $r_i$, the recommendation distribution $(p_i)_{i =1, 2, \dots, N}$ that maximizes entropy, $- \sum_{i=1}^M p_i \ln (p_i)$
subject to an engagement lower, $\E_{r \sim F} [r] \ge a$, is \softmax$(\beta)$ for some $\beta \in \R$.
\end{theorem}

%We show that the optimality of \softmax{} no longer holds in dynamic settings by proposing a simple recommendation algorithm that improves both upon engagement and upon diversity.
We show that the optimality of \softmax{} no longer holds in dynamic settings by proposing an algorithm which deliberately limits availability of content, yet outperformes \softmax{} both in terms of diversity and engagement.

\begin{figure}
    \centering
    \begin{subfigure}[t]{0.4\textwidth}
    \vskip 0pt
    \centering
    \includegraphics[height=2.9cm]{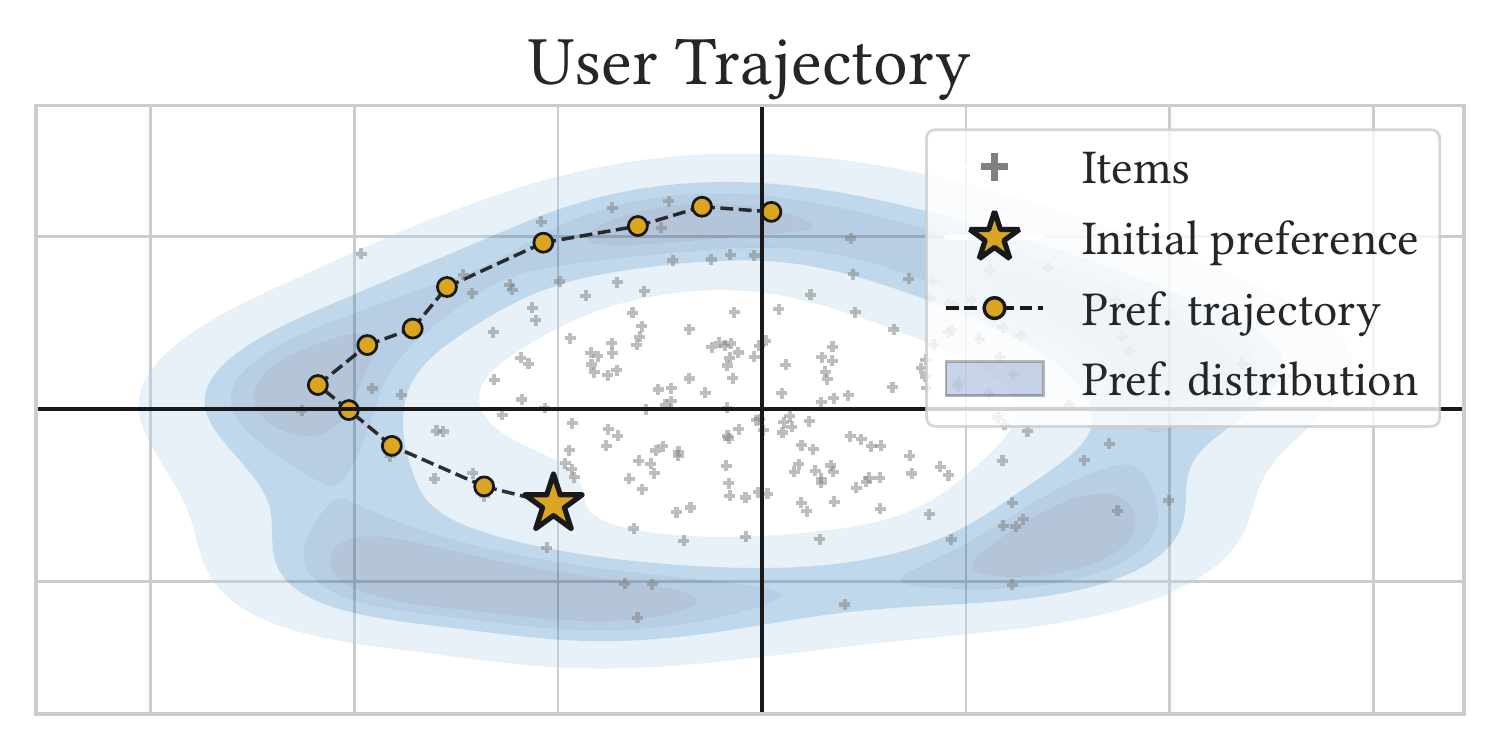}
    \caption{Preference trajectory ($\gamma_{ME} = 0.1$)}
    \end{subfigure}
    \begin{subfigure}[t]{0.59\textwidth}
    \vskip 0pt
    \includegraphics[height=2.9cm]{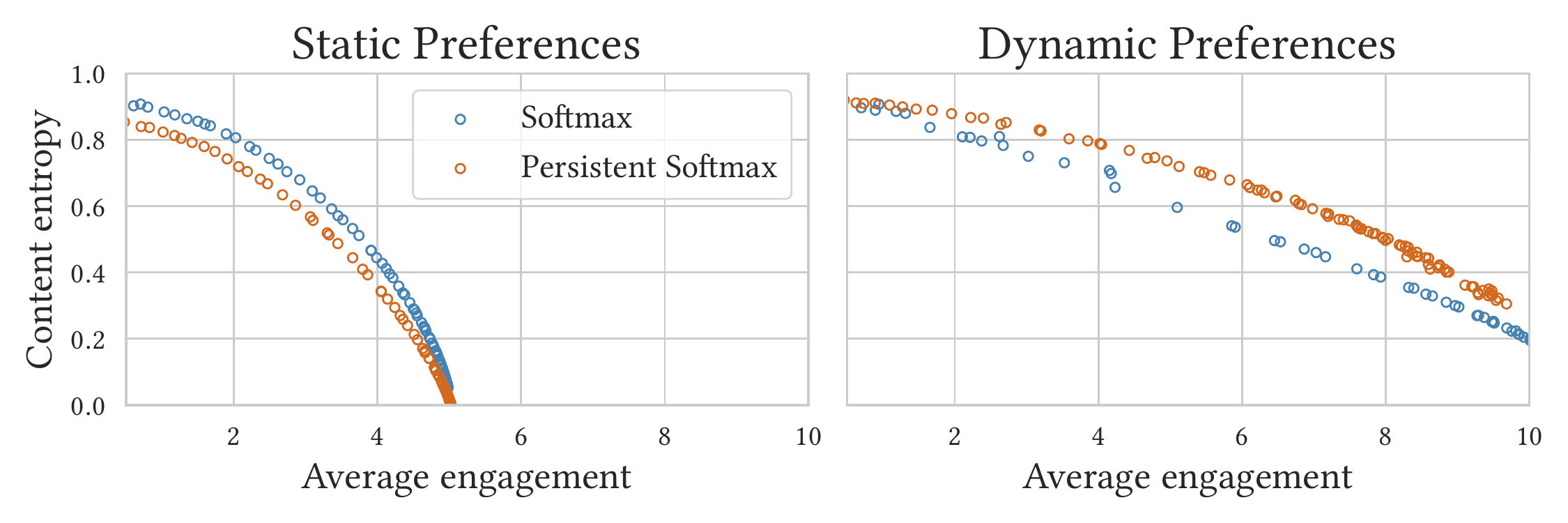}
    \caption{Engagement vs. diversity}
    \end{subfigure}
    \caption{\textbf{\Persistent{} recommendation ($d=2$, $N=1000$):} (a) Preference trajectory with Mere-Exposure dynamic $\gamma_{ME} = 0.1$. The 'clockwise' bias of the recommendations leads to faster 'circling' in preference space. (b) Trade-offs between engagement and diversity for static preferences $(\gamma_{ME} = 0)$ and dynamic preferences $\gamma_{ME} = 0.1$. \Persistent{} is sub-optimal for static preferences but performs strictly better in both content diversity and engagement when preferences are dynamic.}
    \label{fig:biased}
\end{figure}

\paragraph{The \softmax{} with Momentum} As observed in \autoref{subsec:mepred}, Mere-Exposure users tend to circle around, and a recommender might use this behavior to move beyond the Pareto frontier of what is possible with a \softmax{} recommender. \autoref{fig:biased} shows the application of a recommendation policy that \enquote{nudges} user preferences to shift particularly strongly. The recommendation policy used, and benchmarked against the statically optimal \softmax{} policy for different temperatures is
\begin{equation}
\P[v_{t+1} = v] \propto e^{\beta \langle v, u_t \rangle} \ \mathds{1}_{\langle \hat u_t - \hat u_{t-1} , v \rangle > 0},\label{eq:persistent}
\end{equation}
where $\mathds{1}$ is the indicator function. One might call \eqref{eq:persistent} a \emph{persistent} recommender or a \emph{recommender with momentum}, that only recommends content that is in the half of the space that the user moved to in the last two directions. This recommender, by deliberately changing user preferences, allows for higher entropy of the consumed content.
\paragraph{Recommendation System Evaluation with Dynamic Preference Models}
We see that the persistent softmax recommender seems to be preferable to vanilla softmax on both engagement and diversity dimensions. However, this points to a potential gap between the goal of diversity and its operationalization as consumption entropy. The recommendation system deliberately changes users' preferences to increase diversity of consumed content, which arguably is a undesirable property. Hence, metrics for  recommendation systems assuming static users, such as ratings for engagement and entropy for diversity might not capture long-term recommendation system health. In practice, algorithms that seems to improve both in terms of engagement and diversity of consumption might be more opaque than the \enquote{manipulative} recommender considered here, which requires making explicit trade-offs in metrics. 

%%%%%% CONCLUSION.  %%%%%
\section{Avenues for Future Work}\label{sec:conclusions}
% This paper studies psychologically-grounded dynamic preference models for three behavioral effects relevant to recommendation systems: Mere-Exposure, Operant Conditioning and Hedonic Adaptation. Mere-Exposure is the effect that the more familiar something is, the more it is liked.  Operant Conditioning captures the effect that preference for something increase when associated to positive surprise, and decrease when the surprise is negative with respect to a baseline expectation. Lastly, Hedonic Adaptation is the effect that after long enough, all happiness will return to a base level. Next, we formalize quantitative models for these models and find that they have qualitatively different testable predictions. We review psychological and experimental evidence relating to each effect and proposed how to refine said models. We generalize our approach into a broadly applicable methodology for grounding dynamic preference models. We conclude by showing that recommender systems evaluation is impacted by dynamic user models.

We highlight two avenues for future work.

\paragraph{Estimation} This work gives examples of plausible models describing the effect of consumption of content on user preferences and studies the qualitative properties of these models. The statistically efficient and scalable estimation of user models from empirical data is an important step for future work. Here, we distinguish between estimating the strength of posited behavioral effects in benchmark datasets and designing online experimental setups. The case of statistical estimation from historical feedback sequences of users falls under the category of (in the case of Mere-Exposure and Hedonic Adaptation, linear) dynamical systems learning, the  collaborative estimation of item factors might imply significant additional complications. In such cases, the item factor estimates might be biased due to dynamics of the user. Tensor completion \cite{liu2012tensor} and recent work in the econometrics of state dependence \cite{torgovitsky2019nonparametric} are promising directions for such estimation. The contextual factors affecting the size of dynamic effects, e.g. conspicuousness for Mere-Exposure \cite{kihlstrom1987cognitive}, is another important dimension for estimation. The design of online experiments requires either access to real content recommendation systems or careful use of small scale proxy user studies that can test given behavioral hypotheses.

\paragraph{Evaluation and Design} Our discussion in \autoref{sec:design} presented an example where in the presence of dynamic user models a recommender with potentially undesirable properties led to higher engagement and diversity proxies than $\softmax$, which is provably optimally trading off these two metrics for a static user. Evaluating recommendation systems and designing for metrics hence requires to include behavioral models. Further studies that propose and investigate recommendation system metrics for users with dynamic preferences are another area for future research.

%%%%%%% BIBLIOGRAPHY.   %%%%%%%
\bibliographystyle{ACM-Reference-Format}
\bibliography{bibliography}

%%%%%% APPENDIX.  %%%%%%
\pagebreak
\FloatBarrier
\appendix
\section{Omitted Proofs}\label{sec:proofs}
\begin{proof}[Proof of \autoref{prop:bounded}]
Denote $-V = \{-v | v \in V\}$ the negations of item factors. Note that the following statement, which we prove by induction in $t \in \N$, is sufficient to prove boundedness:
\begin{claim}
$\pi_t \in \conv (\{ \pi_0\} \cup V \cup -V)$.
\end{claim} 
For $t=0$, the claim is trivially satisfied. Assume hence that $\pi_t \in \conv (\{ \pi_0\} \cup V \cup -V)$. As for any sets $A, B$, $\conv(\conv(A)) = \conv(A)$ and $A \subseteq B \implies \conv (A) \subseteq \conv (B)$, we have,
\begin{equation}
a \in \conv (A) \text{ and } A \subseteq \conv (B) \implies a \in \conv (B).\label{eq:convisidempotent}
\end{equation}
Applying this statement for $a \coloneqq \pi_{t+1}$, $A \coloneqq \{v_t, \pi_t\}$ and $B \coloneqq ( \{ \pi_0\} \cup V \cup -V)$. $A \subseteq \conv (B)$ holds by the induction hypothesis. $a \in \conv (A)$ holds by the the definition of Mere-Exposure, Operant Conditioning and Hedonic Adaptation, as well as $\surprise \in [-1,1]$. This concludes the induction step. Hence, the claim, and hence boundedness, holds.
\end{proof}
\begin{proof}[Proof of \autoref{thm:boltzmann}]
Consider the following optimization problem:
\begin{align*}
\max_{p_1, p_2, \dots, p_N}& - \sum_{i=1}^N p_i \ln (p_i)\\
&\text{ s.t. } \sum_{i=1}^N p_i = a, \sum_{i=1}^N p_i r_i = 1.
\end{align*}
The corresponding Lagrangian is:
\[
\mathcal L (p, \lambda, \mu) = - \sum_{i=1}^N p_i \ln p_i + \lambda \left(\sum_{i=1}^N p_i - 1\right) + \mu \left( \sum_{i=1}^N p_i r_i - a\right),
\]
The stationarity conditions implies:
\begin{equation}
- \ln p_i - 1 + \lambda + \mu r_i = 0,\label{eq:boltzmann}
\end{equation}
 Rearranging \eqref{eq:boltzmann} yields: 
\[
p_i \propto \exp(-1 + \lambda + \mu r_i).
\]
which is of the softmax form for $\beta = \mu$. The values of $\lambda$ and $\mu$ are such that the feasibility constraints $\sum_{i=1}^N p_i = 1$ and $\sum_{i=1}^N p_i r_i = a$ are satisfied.  
\end{proof}
 
\section{Additional Simulations}\label{app:additional}
In this section, we provide robustness checks for our simulations. There are two kinds of robustness checks. One considers the case of a better estimate of user preferences. As the most drastic example, we consider the \emph{oracle model}, in which the recommender system's estimate is perfect. We also provide a study of a combination of Mere-Exposure and Operant Conditioning, which we did not present in the main text.
\subsection{Perfect Estimation}
\begin{figure*}
    \centering
    \begin{subfigure}[b]{1\textwidth}
         \centering
         \includegraphics[width=\textwidth]{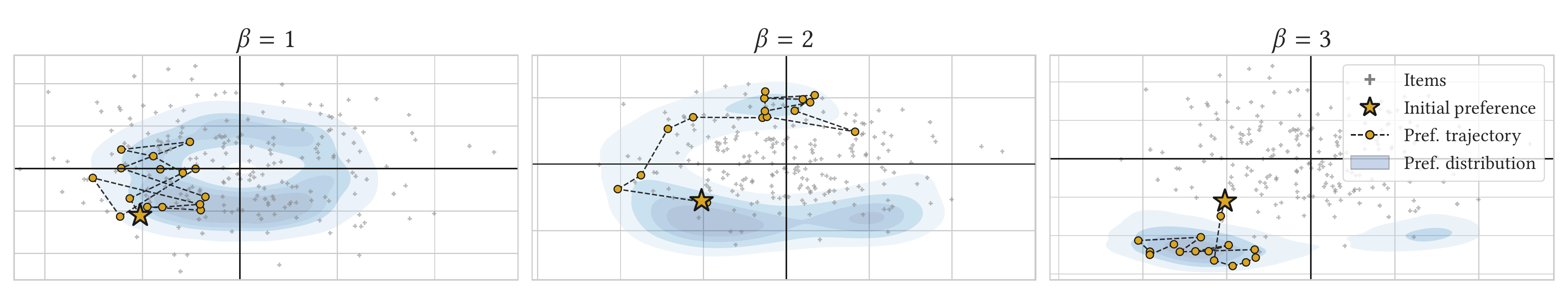}
         \caption{Mere-Exposure}
         \label{fig:me_oracle}
     \end{subfigure}
     \begin{subfigure}[b]{1\textwidth}
         \centering
         \includegraphics[width=\textwidth]{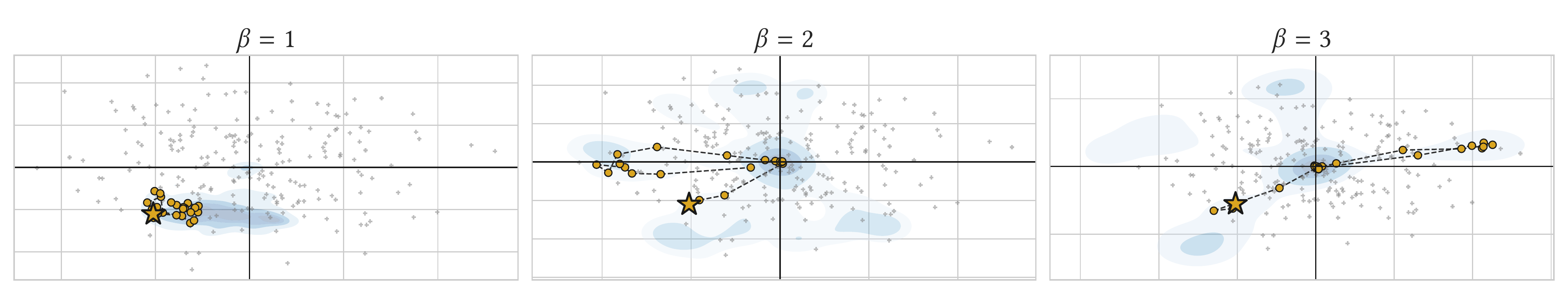}
         \caption{Operant Conditioning}
         \label{fig:oc_oracle}
     \end{subfigure}
         \begin{subfigure}[b]{1\textwidth}
         \centering
         \includegraphics[width=\textwidth]{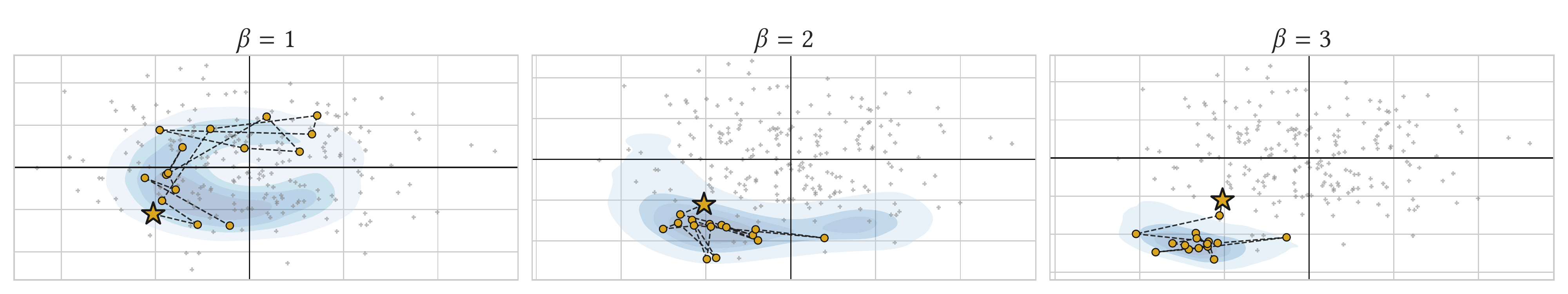}
         \caption{Mere-Exposure and Hedonic Adaptation}
         \label{fig:me_ha_oracle}
     \end{subfigure}
     \begin{subfigure}[b]{1\textwidth}
         \centering
         \includegraphics[width=\textwidth]{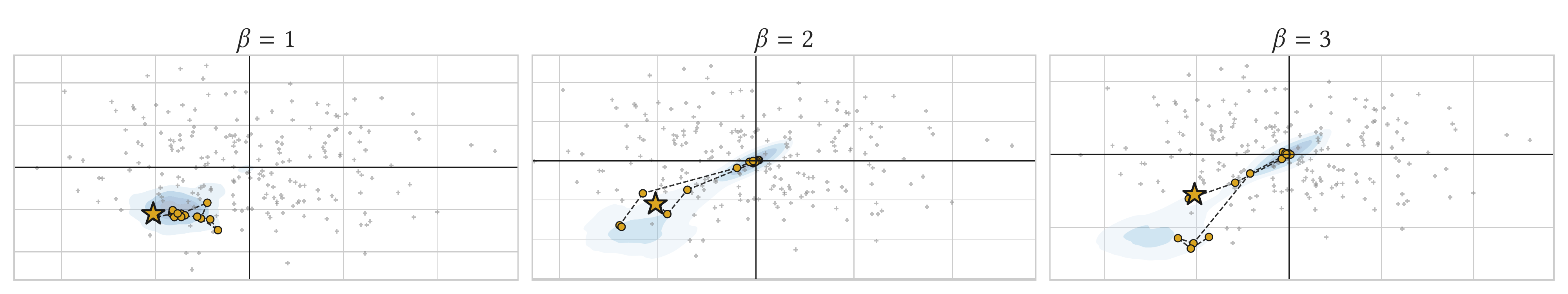}
         \caption{Operant Conditioning and Hedonic Adaptation}
         \label{fig:oc_ha_oracle}
     \end{subfigure}
    \caption{\textbf{Oracle Model:} Preference dynamics with no estimation error}
    \label{fig:oracle}
\end{figure*}

\autoref{fig:oracle} shows the user preference dynamics for no estimation error. The effects we observe mostly align with what was observed earlier: Mere-Exposure dynamics leads to circling in preference space. Where the magnitude of the preference increases with the \softmax temperature parameter and the \enquote{speed} of the dynamic decreases when $\beta$ increases. Operant Conditioning displays similar oscillatory patterns as in the case when recommendations are served based on estimated scores. Mixed dynamics with Hedonic Adaptation show similar patterns, biasing the preference trajectories towards the initial preference.
\subsection{Mere-Exposure and Operant Conditioning}
\begin{figure*}
    \centering
    \begin{subfigure}[b]{1\textwidth}
         \centering
         \includegraphics[width=\textwidth]{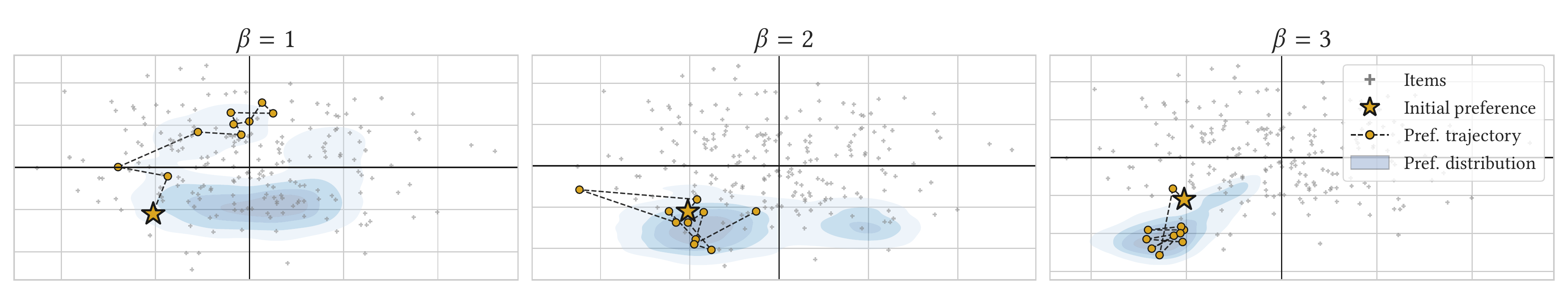}
         \caption{$\gamma_{ME} = \gamma_{OC} = 0.1$}
         \label{fig:strong_me_oc}
     \end{subfigure}
     \begin{subfigure}[b]{1\textwidth}
         \centering
         \includegraphics[width=\textwidth]{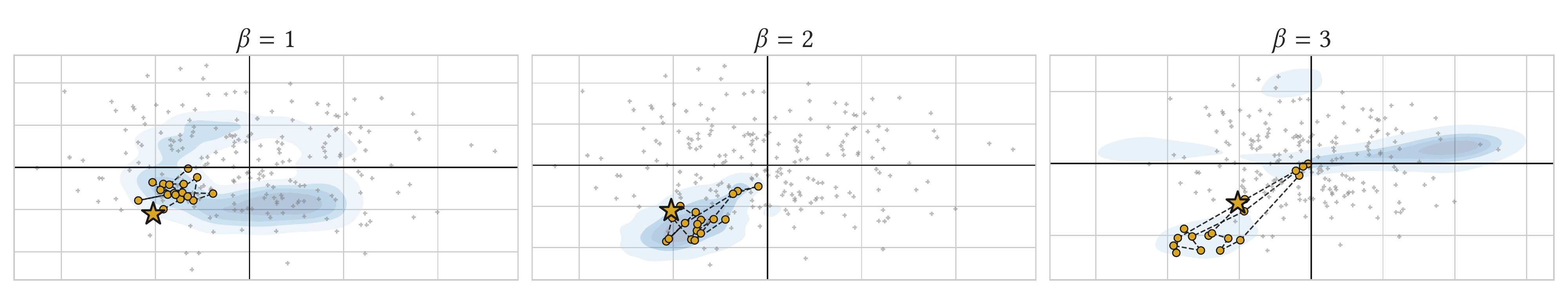}
         \caption{$\gamma_{ME} = 0.05$, $\gamma_{OC} = 0.1$}
         \label{fig:weak_me_oc}
     \end{subfigure}
    \caption{\textbf{Mere-Exposure and Operant Conditioning}: Preference trajectories}
    \label{fig:me_oc}
\end{figure*}

\autoref{fig:me_oc} display the preference trajectory of the combined Mere-Exposure and Operant Conditioning dynamics. This experiments allows us to understand the relative strengths of the $\ME$ and $\OC$ effects. In \autoref{fig:strong_me_oc} the trajectory qualitatively resembles Mere-Exposure. In \autoref{fig:weak_me_oc} we half the strength of the $\ME$ effect which yields a preference trajectory with the oscillatory trademarks of Operant Conditioning.

\end{document}